\documentstyle[12pt,aasms4]{article}

\slugcomment{Submitted to The Astrophysical Journal Letters}

\lefthead{Schiminovich et al.}
\righthead{NUVIEWS Paper I - Continuum}

\begin{document}
\title{The Narrow Band Ultraviolet Imaging Experiment \\
	for  Wide-field Surveys (NUVIEWS)-I: \\
        Dust Scattered Continuum}

\author{David Schiminovich\altaffilmark{1, 2}, Peter Friedman\altaffilmark{1, 2}, Christopher Martin\altaffilmark{2, 1}}


\and

\author{Patrick Morrissey\altaffilmark{2}}



\altaffiltext{1}{Department of Physics, Columbia University, NY, NY, 10027} 
\altaffiltext{2}{Division of Physics, Mathematics, and Astronomy, California Institute of Technology, 405-47, Pasadena, CA 91125} 



\begin{abstract} 

We report on the first results of the Narrow-band Ultraviolet Imaging
Experiment for Wide-field Surveys (NUVIEWS), a sounding rocket
experiment designed to map the far-ultraviolet background in four narrow
bands. This is the first imaging measurement of the UV background to
cover a substantial  fraction of the sky.  The narrow band responses
(145, 155, 161, and 174 nm, 7-10 nm wide) allow us to isolate background
contributions from dust-scattered continuum, H$_2$ fluorescence, and CIV
155 nm emission. In our first flight, we mapped one quarter of the sky
with $\sim$5-10 arcminute  imaging resolution. In this paper, we model
the dominant contribution  of the background, dust-scattered continuum.
Our data base consists of a map of over 10,000 square degrees with 468
independent measurements in 6.25 by 6.25 deg$^2$ bins.  Stars and
instrumental stellar halos are removed from the data. We present a map
of the continuum background obtained in the 174 nm telescope. We use a
model that follows Witt, Friedman, and Sasseen (1997: WFS) to account
for the inhomogeneous radiation field and multiple scattering effects in
clouds. We find that the dust in the diffuse interstellar medium
displays a moderate albedo (a=0.55$\pm$0.1) and highly forward
scattering phase function parameter (g=0.75$\pm$0.1) over a large
fraction of the sky, similar to dust in star forming regions.  We also
have discovered a significant variance from the model.

\end{abstract}




%

\section{Introduction}

Paresce \& Jakobsen (1980) presented the first evidence for a
correlation between the far-UV continuum background and Galactic HI,
thereby determining that most of the background is Galactic in origin,
consistent with scattering of the integrated UV radiation field by
pervasive interstellar dust associated with gas clouds.  This
correlation is typically expressed as a slope of far-UV continuum
intensity vs. neutral hydrogen column $N_{HI}$ (infrared intensity
$I_{FIR}$) and has been measured by many different investigators as
summarized in Bowyer (1991).  ($I_{FUV} \simeq (0.3 $~to~$
2.5)N_{HI}+300$)\footnote{ $N_{HI}$ in units of 10$^{18}$ HI
cm$^{-2}$; $I_{FUV}$ in CU ; 1 Continuum Unit (CU) = 1 photon
cm$^{-2}$ sr$^{-1}$ s$^{-1}$ \AA$^{-1}$} The slope has not yet seen a
true convergence toward a universally accepted value.  This is very
likely due to variations in the dust properties and anisotropies of
the illuminating UV stellar radiation field. Studies of the far-UV
continuum can be used to study the scattering properties of
dust-grains---including positional and spectral dependencies---and
relate them to the physical characteristics of the grains, to probe
the connection between scattering grains and the gaseous components of
the ISM, and to use a synoptic picture of dust-scattering properties
and measured UV background to determine the nature of the interstellar
UV radiation field.

Many of the early investigations into dust-scattering in the
ultraviolet probed the scattering properties of reflection nebulae and
other dark clouds.  Typically, the dust properties extracted from such
studies (in addition to the wavelength-dependent extinction curve),
are the albedo, $a$, and, $g$, the asymmetry parameter of the
conventionally employed Heyney-Greenstein scattering phase
function. For nebular studies, the surface brightness of an externally
illuminated nebula is strongly dependent on $g$ and only weakly
dependent on the albedo $a$ (as one might expect given an anisotropic
incident radiation field), the converse being true for an internally
illuminated nebula.  Dust-scattering of Galactic starlight presents a
more complex geometry, although low-latitude studies (where the
radiation field is relatively isotropic) will better constrain the
albedo, while high latitude studies of predominantly back-scattered
starlight will help determine the asymmetry factor (more typically
$a(1-g)$).

Hurwitz, Bowyer \& Martin (1991; hereafter HBM91) determined the
albedo and scattering asymmetry parameter of interstellar dust from
diffuse background spectra from the UVX experiment.  Unique to the
HBM91 study was a measurement at {\sl low galactic latitudes} that was
relatively uncontaminated by residual starlight. Furthermore, spectra
taken at each point permitted the removal of the contribution from
other components of the background, such as H$_2$ fluorescence and C
IV emission.  Using a plane-parallel, azimuthally symmetric model that
accounted for clumped dust, HBM91 determined a very low albedo,
$0.13<a<0.24$, and nearly isotropic scattering (g$\sim$0). These
values were considerably lower than those previously determined from
diffuse galactic light and nebular studies, as well as lower than
expected from theoretical predictions of grain scattering
(Mathis et al. 1977; Draine \& Lee 1984).

Witt et al (1997; hereafter WFS) employ a Monte-Carlo model of the cloud
structure of the ISM and use far-UV stellar fluxes from the $TD-1$ catalog
(Gondhalekar 1980) to create a two-dimensional radiation field. They
compare the model to measured diffuse background fluxes in 14 fields
obtained with the $FAUST$ experiment (Bowyer et al 1993).  They
determine an albedo of 0.45$\pm$0.05 with an asymmetry factor $g = 0.68
\pm 0.10$.  The dust albedo is 50\% lower than that seen in reflection
nebulae, suggesting a difference between grain sizes (e.g. HII
regions show a deficiency of small grains; Baldwin et al
1991). 

WFS use their model to show that~{\sl the non-isotropic phase function
and radiation fields produce a wide range of I$_{FUV}$/N$_{HI}$ slopes
and could account for the wide variation in measured values.}  The
$FAUST$ measurements produced a high $I_{FUV}/N_{HI}$ slope
($\sim2.5$) CU/10$^{18} cm^{-2}$ because forward scattering grains
couple the bright stars in the plane below to the diffuse radiation
field.  Correspondingly, an isotropic model would underpredict the
albedo in a quadrant deficient of stars, as was the case for the UVX
mission.

We present in this letter the first results from an experiment
specifically designed to spectrally isolate and globally map the
components of the cosmic ultraviolet background.  We focus here on the
dust-scattered continuum and the implications for the properties of the
dust in the diffuse ISM.

\section{Instrument/Flight}

The Narrowband Ultraviolet Imaging Experiment for Wide-Field Surveys
(NUVIEWS) is a rocket-borne instrument that maps line and continuum
components of the far-UV background across 1/4 of the sky
($\sim$10,000$^{\circ^2}$) in 300 s during a single flight. The
experiment contains four co-aligned, wide-field ($20^\circ \times
30^\circ$) self-filtering telescopes.  The three-mirror telescopes
(Hallam, Howell \& Wilson 1990) focus light onto imaging microchannel
plate detectors (Friedman et al 1996).  Self-filtering is achieved
using narrowband all-dielectric multilayer reflective filters (Zukic
et al, 1992) coated directly onto the mirror surfaces.  This design
produces bandwidths of 7-10 nm.  The four telescopes have bands
centered on CIV (155 nm), the peak of the H$_2$ Lyman fluorescence
band (161 nm), and two relatively line-free regions (145 nm) and (174
nm). The experiment was carefully designed to significantly attenuate
the contributions from off-band airglow lines, geocoronal Ly $\alpha$,
earth-shine and moonlight.  All instrument surfaces and baffles
provide the maximum possible stray and scattered light rejection,
given the constraints of a wide-field mission.  This paper
concentrates exclusively on data from the fourth band (174 nm) which
is a region relatively free of molecular and atomic transition lines,
ideal for studying dust-scattered far-UV continuum light. Other
instrument parameters are summarized in Table 1.

The payload was launched at 07:15:00.7 (UT) on July, 14 1996. 
NUVIEWS implemented a zig-zag pattern consisting of eight
scans of 90$^\circ$ in length (measured using the field center) with the
whole rectangular scan area centered on R.A. = 19 h, Dec = 30$^\odot$. 
All data presented in this paper were acquired during the 287 seconds
that the payload was above 150 km altitude.

\section{Calibration and Data Analysis}

The $TD-1$ database of far-ultraviolet (156.5 and 197.5 nm) stellar
photometry (Gondhalekar 1980) is the ideal catalog for comparison with
our images, since the $TD-1$ flux limit (5.0$\times 10^{-13}$ erg
cm$^{-2}$ \AA$^{-1}$ s$^{-1}$) is comparable to our detection limit
and we can ascertain a position for nearly every star that we
detect. The $TD-1$ flux scale is known to be non-linear below fluxes
of 1.0$\times 10^{-12}$ erg cm$^{-2}$ \AA$^{-1}$ s$^{-1}$ (Bowyer et al
1993).  We incorporated the higher flux limit when cross calibrating
stars during the in-flight calibration.

After applying corrections for telescope plate scale, distortions and
offsets, detector non-linearities, position offsets, pulse-height
dependencies, rotations and thermal drift, we used the positions from
the TD-1 stellar database to recover the telescope aspect
solution. The solution yielded 8$^\prime$-10$^\prime$ FWHM for the 174
nm telescope, nearly identical to the pre-flight performance.
Superflat, star-subtracted sky images were generated in detector
coordinates using the complete set of scan data.  Normalized,
linearity-corrected superflats were combined with the aspect solution
to generate exposure sky-maps corresponding to the instrument
area-bandwidth-exposure product.  The absolute calibration of these
maps was refined based on in-flight measurements of TD-1 stars.

We then generated exposure-corrected 20$^\circ \times 20^\circ$ field
maps on 15$^\circ$ centers. When studying the background, extended
point spread function halos can produce a spurious signal,
particularly in regions around bright stars.  During post-calibration
we characterized the shape of the point spread function out to
5$^\circ$ radius.  This shape was also compared with the in-flight
data using stars with no apparent dust halo.  Simulated maps of the
TD-1 fields convolved with the telescope PSF were generated for each
field and subtracted from the calibrated data maps.  This allowed us
to accurately subtract the flux due to the stellar component as well
as to verify the instrument calibration. Since some detected stars are
not present in the TD-1 catalog, and the subtraction at the peak of
the PSF is not perfect, we blanked ($\sim$1$^\circ$ diameter) around
TD-1 star locations and other peak detections.  The remaining unmasked
flux was binned into square bins of 3.75$^\circ$, 5$^\circ$, and
6.25$^\circ$ on a side.  Heavily masked bins, or bins with low
exposure (e.g. at the edge of the scan pattern) were flagged and
omitted from the subsequent analysis. The 6.25$^\circ$ binned data, in
468 independent bins, were used for comparison with our
dust-scattering model, described in the next section.  We obtained
values for N$_{HI}$ and I$_{IR}$ for each bin, only using data from
the unmasked regions in our field maps. (Composite HI maps from Dickey
\& Lockman, 1990 and 100 $\mu$m IR maps from Schlegel et al 1998).  We
plot I$_{FUV}$ vs. N$_{HI}$ in Figure 1, where a significant
correlation is apparent.

We also constructed a dust-scattered continuum sky-map, shown in
Figure 2.  In generating this image, we smoothly interpolated over
masked regions using a $\sim1.5^\circ$ smoothing diameter.  The sky-map
reveals dust-scattered halos around bright stars (e.g. near the Upper
Scorpius reflection nebula at l=0$^\circ$, b=20$^\circ$) and a strong
gradient decreasing upwards from the plane.  Also evident is a clear
variation with Galactic longitude.  There exists a significant
depression in dust-scattered continuum light at high latitudes above
less intense segments of the Galactic plane from l=20$^\circ$ to
l=60$^\circ$.  

\section{Dust Scattering Model}

We have duplicated the radiative transfer model of WFS to model the
continuum data. The model uses the TD-1 catalog to produce a (2D)
radiation field on the sky. All stars are assumed to be at infinity. The
model then calculates surface brightnesses for three types of cloud
illuminated by this radiation field, including the effects of multiple
scattering, for a grid of albedos and phase function parameters (a,g).
Finally, for each galactic (l,b) point in our map, and the corresponding
N$_{HI}$, the model calculates (statistically)
the distribution of cloud numbers and types that produces this N$_{HI}$ and
the corresponding scattered intensity. Several hundred realizations are
calculated for each line of sight to obtain an average intensity. The
model makes a definite prediction for the intensity at every point in
the map.

We used two versions of this model. Model 1 uses the cloud
distribution of WFS, with three types having central optical depths
$\tau_{1,2,3}=$0.5, 2.5, and 10, and frequency ratios 12:3:1.  The
more opaque clouds show significant multiple scattering effects, and
reflect less efficiently than the same material spread out over lower
densities.  

We have two concerns with this approach. Firstly, the distribution
chosen by WFS we feel underestimates the contribution of small, low
optical depth clouds which dominate the cloud number distribution
(Kulkarni \& Heiles 1987). Secondly, even the correct model for HI
absorption misses the Warm Neutral Medium (WNM) traced by HI emission
that is weakly absorbing. The WNM accounts for 25-50\% of the HI in the
local Milky Way (Kulkarni \& Heiles 1987; Braun 1998) Thus we also
examined a model (2) with a fourth, low optical depth cloud
($\tau_4=$0.05), where cloud types 1:2:3:4 have a frequency distribution
12:3:1:76. Model 2 produces about 14-17\% more UV flux at low HI column
densities, and 2-5\% more at high HI column densities.

When we compare the model and data, we perform several data cuts.  We
only use data with exposure times exceeding 10 seconds, bins that have
more than 20\% of their area remaining after star blanking, and no
bins at the edges of the map.  We excluded two regions with clear
stellar halos and three anomalously low bins.

A chi-square comparison of model and data show a significant
additional source of variance over that predicted from counting
statistics. We have verified that this is not correlated with the
brightness of stars that are removed or star halos that are subtracted
from the maps. We find no evidence for any other variance introduced
by the instrument or analysis.  The variance appears to increase
linearly with the intensity, and correspondingly the fractional
deviation decreases with the square root of the intensity or
N$_{HI}$. This is consistent with a source of variance produced along the
line of sight which due to the accumulation of more samples would grow
linearly with the total number of sources.  In order to obtain a
sensible chi-square statistic to obtain confidence limits, we model
this ``cosmic variance'' as ${\sigma_I}^2=aI$. The value $a=25$ CU provides a
good fit to the variance rise with I in CU. For example, at the
lowest intensities (I=400 CU) the rms deviation from the model is
about 25\%.

Even with this substantially increased variance, a chi-square
comparison yields a tight limit on acceptable (a,g) pairs. The model
includes an extragalactic background component of 300 CU and the
appropriate absorption in the galaxy. We also minimize chi-square by
adding an additional unknown flux to the model, either as a simple
constant, or as an airglow flux that is proportional to the slant
airmass through the residual atmosphere.  We considered several
minimum latitude cuts ($b_{min}<|b|$), which had a modest
effect. Slant component typical values are a=0.55$\pm$0.1 and
g=0.75$\pm$0.1. We also calculated the slope of the data vs. N$_{HI}$,
IRAS 100 $\mu$m, and csc b, compared with the various models. The best
fit models reproduce all three slopes well. An offset over and above
the 300 CU extragalactic component of 150$\pm$50 CU is required to
produce the best fits.  Model calculations indicate that a small
component of the residual ($<$10\%) may be due to unresolved starlight
that has not been masked or subtracted out. A summary is presented in
Table 2 and the data and best fit model are compared in Figure 3.

\section{Discussion}

The NUVIEWS survey is unique in combining significant sky coverage over
a contiguous region, with good imaging and star removal. It provides the
first global measure of the properties of the dust-scattered continuum,
while also measuring local variations with significance.  We summarize
here the implications for the continuum-HI slope, dust scattering
properties, the variance of the correlation residuals and the offset
in the continuum-HI correlation.

Our measured slope (0.35-0.50) is among the lowest slopes yet measured, but
is consistent with the weaker stellar radiation field observed at these
longitudes.  We thus confirm the prediction of WFS that the slope can
vary by more than a factor of five due to the anisotropic radiation
field.

The best model fit values for albedo and phase function asymmetry
(a=0.55$\pm0.1$, g=0.80$\pm0.1$) are intermediate between the values
measured from the diffuse radiation field by WFS and from  dust
scattering in star forming regions (Calzetti et al 1995; Witt et al
1992) and are consistent with both within all the uncertainties.  We
find no evidence that dust scattering properties are not universal in
disparate environments.

Local variations are seen quite clearly in the width of the
I$_{UV}$-N$_{HI}$ relation, and the variance in the best model fit,
illustrated in Figures 1 and 3.  Possible origins are the 3-D inhomogeneity
of the radiation field, variance in cloud properties, and local
fluctuations in the dust scattering properties.  Further analysis, for
example by correlation of the residuals with other observables, may
clarify the source of the variance.

Finally, the origin of the model continuum offset, and its relation to
the other phases of the ISM (for example the Warm Ionized Medium)
remains an open question.  This and other open issues raised in this
paper will be addressed in future analyses of this data set, corollary
data sets and the balance of the NUVIEWS all-sky survey.

\nocite{KuHe87}
\nocite{ScFi99}
\nocite{Br98}
\nocite{BoSa93}
\nocite{PaJa99}
\nocite{GoPh80}
\nocite{WiFr99}
\nocite{mat90}
\nocite{WiPe92}
\nocite{CaBo95}
\nocite{Bo90}
\nocite{HuBo91}
\nocite{MaRu77}
\nocite{DrLe84}
\nocite{FrCu96}
\nocite{ZuTo92}
\nocite{HaHo83}
\nocite{DiLo90}

\acknowledgments

We would like to gratefully acknowledge Steven Kaye, Irwin Rochwarger,
Judy Fleischman, Peter Marshall, Muamer Zukic, and NASA project manager
David Kotsifakis and his team for work on the sounding rocket payload.
This work was supported by NASA grant NAG 5-642 and NAG 5-5052 and
funding from the California Institute of Technology.

\clearpage

\clearpage

\figcaption[fig1.ps]{Intensity in the 174 nm map (in CU) vs. N$_{HI}$
column density (in units of 10$^{20}$ cm$^{-2}$), for
$b_{min}=22.5^\circ$. Error bars include ``cosmic variance''
component. Best fit is line with slope 0.42 CU/10$^{18}$cm$^{-2}$,
offset 530 CU.  }

\figcaption[fig2.ps]{NUVIEWS 174 nm calibrated sky-maps using flight 1
data. Aitoff projections are centered on Galactic center.  {\it Main:}
Star-subtracted continuum map.  Color bar indicates intensity in
CU. Pixel size is 1$^\circ \times 1^\circ$.  {\it Inset:} Sky-map with
stars included.  Color is scaled to the cube-root of the intensity, in
arbitrary units. Pixel size is 0.5$^\circ \times 0.5^\circ$.}

\figcaption[fig3.ps]{Intensity in the 174 nm map (in CU) vs. model
2a, for $b_{min}=22.5^\circ$.  Error bars include ``cosmic variance''
component.  }

\clearpage

\singlespace
\begin{deluxetable}{lc}
\tablecolumns{2}
\footnotesize
\tablecaption{NUVIEWS Instrument Summary (174 nm Telescope)} \label{tab:instover}
\tablewidth{0pt}
\tablehead{
\colhead{Parameter} & \colhead{Value} }
\startdata
{\bf Telescope} & Three mirror, Wide-Angle, Flat-Field (WAFFT)\nl
Focal Length & 90 mm \nl
Focal Ratio & 3.0 \nl
Field of View (Unobscured)& $30^{\circ} \times 20^{\circ}$ \nl
Angular Resolution & $5^\prime-10^\prime$ \nl
Filter Location & Primary, Secondary, Tertiary Surfaces \nl
Bandwidth & 10 nm \nl
{\bf Detector} & 70 mm Microchannel Plate \nl
Photocathode & CsI \nl
X, Y Position Readout & Delay-Line Anode \nl 
Detector Window & MgF$_2$ \nl
Detector Filter & Fused Silica \nl
{\bf Effective Area$\times$Bandwidth} & 1.46  cm \AA \nl
{\bf Average Exposure} & 15 s \nl

\enddata
\end{deluxetable}

\begin{footnotesize}
\begin{table}
\begin{tabular}{ccccccccccc} \\ 
\multicolumn{11}{l}{Table 2 -- Continuum Model Fit Summary} \\ \hline
b$_{min}$ & Model$^a$ & a & g & $\chi^2$$^b$ & dof & 
I$_0$ & c$_{HI}$[model]$^c$ & c$_{HI}$[data]$^c$ &
c$_{d}$[model]$^d$ & c$_{d}$[data]$^d$ \\ \hline
15.0 & 1 & 0.55 & 0.8 & 393  & 242 & 330 & 0.48 & 0.49 & 1.12 & 1.13 \\
22.5 & 1 & 0.55 & 0.8 & 160 & 174 & 320 & 0.51 & 0.42 & 0.92 & 0.91 \\
30.0 & 1 & 0.50 & 0.77  & 139  & 145 & 190 & 0.37 & 0.34 & 0.75 & 0.73 \\
15.0 & 2 & 0.55 & 0.8  & 406  & 242 & 160 & 0.48 & 0.49 & 1.13 & 1.13 \\
22.5 & 2 & 0.60 & 0.77 	& 141 &  174 & 180 & 0.43 & 0.42 & 0.91 & 0.90 \\
30.0 & 2 & 0.60 & 0.77  & 120 &  145  & 180 & 0.41 & 0.35 & 0.75 & 0.74 \\
\end{tabular}

\begin{scriptsize}
\noindent{$^a$ Model 1-WFS; Model 2-WFS plus WNM.}
\noindent{$^b$ Includes cosmic variance with $\sigma_I^2=25 I$ (CU).} 
\noindent{$^c$ Units CU/10$^{18}$ cm$^{-2}$ (HI from Dickey \& Lockman 1990).} 
\noindent{$^d$ Units CU/(MJy/sr) (100 $\mu$m IR from Schlegel et al 1998).} 
\end{scriptsize}
\end{table}
\end{footnotesize}

\clearpage

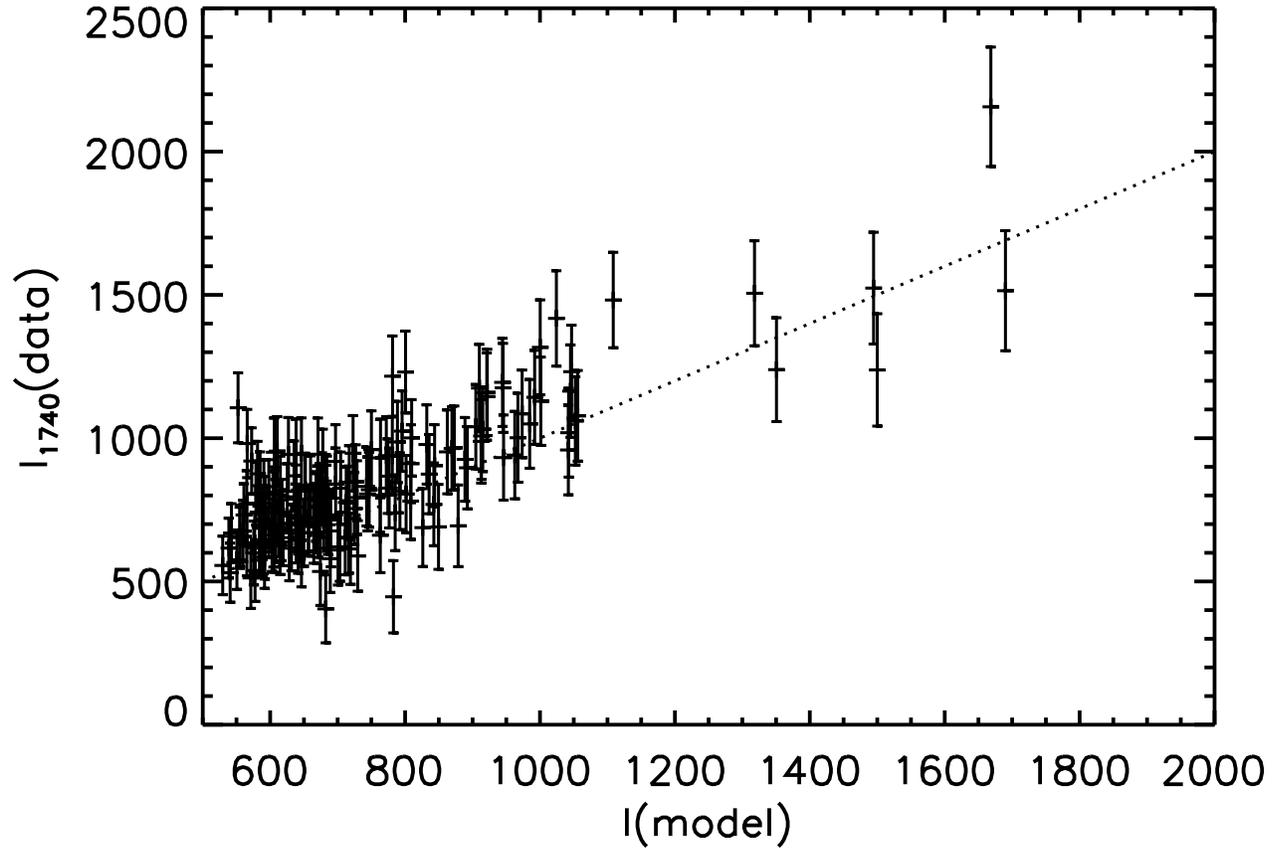
\end{document}